\documentclass{aa}  
\usepackage{graphicx}
\usepackage{txfonts}
\usepackage{dcolumn}
\newcommand{\HH}  {\mbox{H$_2$}}       %  H2
\newcommand{\icm}{\text{cm}^{-1}}
\newcolumntype{x}[1]{D{.}{.}{#1}}
\begin{document}

   \title{The full infrared spectrum of molecular hydrogen
  }

   \author{E. Roueff \inst{1}
          \and
          H. Abgrall \inst{1}%\fnmsep\thanks{Just to show the usage
          %of the elements in the author field}
                \and
          P. Czachorowski\inst{2}
             \and
          K. Pachucki\inst{2}
                \and
          M. Puchalski\inst{3}
             \and
          J. Komasa\inst{3}
    }

   \institute{Sorbonne Universit\'e, Observatoire de Paris, Universit\'e PSL, CNRS, LERMA, F-92190,  Meudon, France \\
              \email{evelyne.roueff@obspm.fr}
       %  \and
       %      University of Alexandria, Department of Geography, ...\\
       %      \email{c.ptolemy@hipparch.uheaven.space}
      %        \thanks{The university of heaven temporarily does not
       %              accept e-mails}
     \and
     Faculty of Physics, University of Warsaw, Pasteura 5, 02-093 Warsaw, Poland \\
     \email{krp@fuw.edu.pl}
     \and
     Faculty of Chemistry, Adam Mickiewicz University, Umultowska 89b, 61-614 Pozna{\'n}, Poland \\
     \email{komasa@man.poznan.pl}
             }

   \date{}

% \abstract{}{}{}{}{} 
% 5 {} token are mandatory
 
  \abstract
   {The high spectral resolution R $\sim$ 45,000 provided by IGRINS (Immersion Grating INfrared Spectrometer) at MacDonald Observatory
  and  R $\sim$ 100,000 achieved by CRIRES (CRyogenic high-resolution InfraRed Echelle Spectrograph) at VLT (Very Large Telescope) challenges the present knowledge of infrared spectra. }
   {We aim to predict the full infrared spectrum of molecular hydrogen at a comparable accuracy.}
   {We take advantage of the recent theoretical {\it{ab initio}} studies on molecular hydrogen to compute both the electric quadrupole and magnetic dipole transitions taking place within the ground electronic molecular state of hydrogen.}
   {We computed the full infrared spectrum of molecular hydrogen at an unprecedented accuracy and derive for the first time the emission probabilities including both electric quadrupole ($\Delta  J = 0, \pm$2) and magnetic dipole transitions  ($\Delta  J = 0$) as well as the total radiative lifetime of each rovibrational state. Inclusion of magnetic dipole transitions increases the emission probabilities by factors of a few for highly excited rotational levels, which occur in the 3-20 $\mu$ range}
   {}


   \maketitle
%
%-------------------------------------------------------------------

\section{Introduction}

  Molecular hydrogen (H$_2$) is a symmetric molecule, which prohibits electric dipole transitions to occur within its  X$^1\Sigma^+_g$ ground electronic state. 
  Then electric quadrupole and magnetic dipole transitions can take place, reciprocally submitted to $\Delta J= 0, \pm 2$ and $\Delta J= 0$ selection rules. 
  Considerable attention has been paid to the ab initio studies of this simple molecule  where the resolution of the one dimensional Schr{\"o}dinger equation allows the energy spectrum of the ground electronic state to be derived, as described in the pioneering work by \cite{kolos:64}. Various corrections have been further introduced in order to compare the theoretical {\it{ab initio}} values to the experimentally derived values by \cite{dabrowski:84} from the VUV (Vacuum Ultra-Violet) absorption and emission flash discharge spectra of H$_2$ in the Lyman and Werner bands. The rotation-vibration energy level measurements were derived for all vibrational levels $v$~=~0-14, and a maximum rotational $J$ value of $J$ = 29 with a predicted accuracy of 0.1 cm$^{-1}$. These measurements challenged theoretical calculations as the discrepancy between experimentally derived values and theoretical calculations reached about 4~cm$^{-1}$ for highly excited rotational levels. An additional important step in the ground level energy determinations was provided by the study of the quadrupole infrared spectrum at the laboratory \citep{bragg:82,jennings:83}, which provides level energy terms with an accuracy of about 0.001 cm$^{-1}$.  These studies, which are limited to energy terms of low vibrational and rotational values, have prompted detailed fundamental ab-initio studies of the H$_2$ spectrum where non-adiabatic, relativistic, quantum electrodynamic (QED) corrections are carefully introduced by \cite{pachucki:09} and \cite{komasa:11}. 
In parallel, astrophysical observations in the infrared have allowed the detection of tens of  highly H$_2$ excited emission rovibrational transitions occuring in so-called Photon dominated regions (PDRs) such as the Orion Bar \citep{kaplan:17} or NGC 7023 \citep{le:17} or in shocked regions such as the Orion KL outflow  \citep{oh:16,geballe:17} and Herbig-Haro objects \citep{pike:16}. Forthcoming infrared facilities, such as CRIRES (CRyogenic high-resolution InfraRed Echelle Spectrograph) at VLT (Very Large Telescope) and JWST (James Webb Space Telescope) in space will allow the wavelength window and the sensitivity to be extended. We find that it is thus timely to provide the complete infrared spectrum of H$_2$  involving any possible transition linking all  available rovibrational levels at the highest level of accuracy.

%--------------------------------------------------------------------
\section{Infrared spectrum of ground state \HH}
\label{sec:1}

\subsection{Quantum mechanical calculations of line positions}
\label{Sec:lp}
Individual energy levels and their wave functions were determined in the framework 
of the nonrelativistic quantum electrodynamics (NRQED) \citep{Caswell:86,Pachucki:05}. 
This theory is suitable for bound energy levels of atomic and molecular systems 
composed of light nuclei. Within this approach, the energy is expressed 
in the form of an expansion in powers of the fine structure constant $\alpha$ 
\begin{equation}\label{alphaexp}
 E(\alpha) = E^{(2)} + E^{(4)} + E^{(5)} + E^{(6)} + E^{(7)} + \mathcal{O}(\alpha^8) \,,
\end{equation}
where $E^{(i)}$ is proportional to $\alpha^i$ and may contain powers of $\ln\alpha$.
The subsequent terms of this expansion are commonly known as 
the nonrelativistic energy $E^{(2)}$, the relativistic $E^{(4)}$, 
the QED $E^{(5)}$, and higher order $E^{(i)},\,i>5$, corrections.
Each term, in turn, can be expanded in the small electron-to-proton mass ratio 
$m_\mathrm{e}/M_\mathrm{P}$ as implemented
in the nonadiabatic perturbation theory (NAPT) by \cite{PK08b,PK:15}.
The nonrelativistic energy $E^{(2)}$ is then obtained by solving the nonadiabatic radial
Schr{\"o}dinger equation composed of the Born-Oppenheimer, 
the adiabatic correction, and the nonadiabatic correction potentials \citep{PK:09}.
For \HH\ these electronic potentials are known to the relative accuracy of $10^{-9}-10^{-14}$
\citep{Pachucki:10b,PK:14,PK:15}, which enables the nonrelativistic
dissociation energy of an individual level to be determined with the accuracy limited
only by the missing higher order nonadiabatic corrections $\sim\!(m_\mathrm{e}/M_\mathrm{P})^3$. 
The latter have recently been found to be of the order of $10^{-4}-10^{-5}\,\icm$ \citep{PK:18a}.

The hydrogen molecule in its electronic ground state (X$^1\Sigma^+_g$)
accommodates 302 bound states\footnote{The $v$'=14, $J$'=4 level is considered as the highest bound level.}.
An apparent advantage of the NAPT approach is that once the necessary
potentials are constructed all the energy levels can be obtained simultaneously.
An alternative to the NAPT and significantly more accurate approach
to the nonrelativistic energy has recently been developed by \cite{PK:18a}. 
In this approach, the $E^{(2)}$ of a selected energy level is evaluated directly 
in a four-body calculation without separation of electronic and nuclear movements.
The accuracy of such calculations is limited by the current precision of the physical
constants and amounts to $10^{-7}\,\icm$ for the dissociation energy of an individual level.
At present, for about one-fifth of all the bound levels much more accurate
nonrelativistic energies are available.

The NAPT was also applied to evaluate the relativistic correction $E^{(4)}$.
The electronic relativistic potential for \HH\ was determined with accuracy
higher than $10^{-6}\,\icm$ by \cite{PKP:17}. Recently, \cite{CPKP:18}
evaluated the leading order recoil $\sim\!m_\mathrm{e}/M_\mathrm{P}$ correction
to this potential so that currently its accuracy is limited by the unknown
higher order terms $\sim\!\left(m_\mathrm{e}/M_\mathrm{P}\right)^2$
and is of the order of $10^{-6}\,\icm$. The uncertainty introduced by the $E^{(4)}$
term to the dissociation energy of a level is estimated as $10^{-5}\,\icm$.

The QED term $E^{(5)}$ was obtained from the electronic QED potential
determined by \cite{PLPKPJ09} and later refined by \cite{PKCP:16} and \cite{PKP:17}. 
So far, no numerical values for the finite nuclear mass (i.e. recoil) correction
to the QED potential are known. This missing correction currently limits the overall accuracy
of the molecular levels energy predicted by theory. The uncertainty resulting from
this missing contribution is estimated as $2\,E^{(5)}/M_\mathrm{P}\sim\!10^{-4}\,\icm$.

The higher order $E^{(6)}$ QED correction was evaluated in the non-recoil limit
by \cite{PKCP:16}. The accuracy of this correction is limited by
two factors that contribute errors of the same order---the unknown 
finite-nuclear-mass corrections and the numerical convergence.
The final uncertainty on the energy of a level is estimated
as $10^{-6}\,\icm$ or less.

Complete expressions for the higher order corrections $E^{(i)},\,i>6$, 
are unknown. For this reason, their numerical values were estimated by \cite{CPKP:18}, 
on the basis of the dominating terms selected by analogy to atomic hydrogen, 
with relatively large uncertainties. However these corrections, as well as the other 
tiny contributions like the finite nuclear size correction, contribute at the level 
of $10^{-4}\,\icm$ or less to the total dissociation energy. 

As mentioned above, the final accuracy of a single energy level is currently restricted by 
the accuracy of QED contribution and changes slowly from state to state.
There is, however, a significant cancellation of the errors carried by individual levels
involved in a transition energy. This is particularly pronounced for near-lying levels.
This cancellation was controlled for each energy contribution separately. 
The final uncertainty determined this way accompanies each line position (see $\Delta\sigma$ column in Table~\ref{tab:2}).
The above uncertainty discussion is illustrated by the numerical data 
displayed in Table~\ref{T:S1H2} concerning the 1-0 S(1) transition.
More details on the theoretical and computational procedures outlined above can be found in
\cite{Komasa:19}.

\begin{table}
\caption{\label{T:S1H2}
Calculated contributions (and  uncertainties) to  S(1) transition energy 
in  \HH\ fundamental band. Values are given in $\icm$.
}
\begin{tabular}{l@{\hspace{16pt}}x{5.11}}
Contribution & \multicolumn{1}{c}{1-0 S(1)} \\
\hline
$E^{(2)}$  		& 4\,712.895\,971\,8(1)  \\
$E^{(4)}$			&      0.034\,872\,6(2)    \\
$E^{(5)}$			&     -0.025\,99(3) \\
$E^{(6)}$ 			&     -0.000\,231\,9(7)    \\		
$E^{(7)}$			&      0.000\,013(3)    \\	
%$E_\mathrm{FS}$	    &     -0.000\,00	   \\
Total				& 4\,712.904\,64(3)  \\
\end{tabular}
\tablefoot{1-0 S(1) corresponds to $v_u$=1, $J_u$=3, $v_l$=0, $J_l$=1, where $v_u, J_u; v_l, J_l$ are the vibrational and rotational quantum numbers of the upper ($u$) and lower ($l$) levels of the transition.}
\end{table}

Both electric quadrupole and magnetic dipole transitions may occur within the X electronic ground state. 
Below we discuss separately both types of the transitions and supply working formulas for them.

\subsection{Electric quadrupole  transitions}
Quadrupole  vibration-rotation transition probabilities of molecular hydrogen were first computed by \cite{turner:77} for vibrational levels up to $v$=14 and rotational levels up to $J$=20. These calculations were subsequently improved and extended by \cite{wolniewicz:98} for all available bound rovibrational levels of H$_2$ with a more accurate quadrupole moment function $Q(r)$.
The electric quadrupole emission probability $W_{v'J' \rightarrow v"J"}$ in s$^{-1}$ is formulated as given in theoretical textbooks (e.g., \cite{sobelman:06}):
\begin{align}
W_{v'J' \rightarrow v"J"}&= \frac{ \alpha }{15}\cdot \frac{\omega^5}{c^4} \cdot \frac{1}{(2J'+1)} \cdot\\ &\sum_{M'M"} \left|\left<J'M' f_{v',J'}(r)\left|r^2 P_2(\cos\theta)\right| J"M" f_{v",J"}(r)\right>\right|_{CGS}^2 \,,\nonumber
\end{align}
where all quantities are expressed in CGS units. $f_{v,J}(r)$ is the radial wavefunction of the ground state \HH~molecule, solution of the radial Schr{\"o}dinger equation, corresponding to the discrete eigenvalue $E_{v,J}$. The symbols $J, M$ stand for the $Y_J^M$ spherical harmonics, solution of the angular part of the Schr{\"o}dinger equation. The $\omega$ is the angular frequency of the transition. 
Introducing the wavenumber $\sigma$ = $\frac{\omega}{2 \pi c}$ and numerical values for 
the physical constants, the previous expression becomes 
\begin{align}
W_{v'J' \rightarrow v"J"}&= 1.4286\times 10^{11}  \cdot \sigma^5  \cdot \frac{1}{(2J'+1)} \cdot \\\nonumber
&\sum_{M'M"} \left|\left<J'M'  f_{v',J'}(r) \left|r^2 P_2(\cos\theta)\right| J"M" f_{v",J"}(r)\right>\right|_{CGS}^2.
\end{align}
The sum over the various substates $M'$, $M"$  gives rise to the $f(J', J")$ angular coeffcients, as reported in \cite{wolniewicz:98}, with specific selection rules corresponding to O ($\Delta J  = -2$), Q  ($\Delta J = 0$) and S ($\Delta J = 2$) transitions.

To evaluate the electric quadrupole moments we employed the radial function (in a.u.)
\begin{equation}
Q(r)=\frac{r^2}{2}-\frac{1}{2}\langle\phi|{\sum_a r_a^2(3\cos^2\theta_a-1)}|\phi\rangle
\end{equation}
with the expectation value evaluated with the Born-Oppenheimer wave function $\phi$.
The calculated values of $Q(r)$, employed already in \cite{pachucki:11} and in \cite{Campargue:12}, 
are in agreement with those obtained by \cite{wolniewicz:98} 
except that the latter represent twice the $Q(r)$.
The working equation used for computing the emission probabilities is
\begin{align}
W_{v'J' \rightarrow v"J"}&= {1.4286\times 10^{11}}  \cdot  a_0^4 \cdot \sigma^5  \cdot \frac{1}{(2J'+1)} \cdot\\\nonumber 
&\sum_{MM'}\left|\left<JM  f_{v',J'}(r) \left|Q(r)\right| J'M' f_{v",J"}(r)\right>\right|^2\,,
\end{align}
\begin{align}
W_{v'J' \rightarrow v"J"}&= {1.12 \times 10^{-22}}  \cdot \sigma^5  \cdot \frac{1}{(2J'+1)} \cdot\\\nonumber 
&\sum_{MM'}\left|\left<JM  f_{v',J'}(r) \left|Q(r)\right| J'M' f_{v",J"}(r)\right>\right|^2\,,
\end{align}
where the wavenumbers $\sigma$ are expressed in reciprocal centimeters (cm$^{-1}$).

\subsection{Magnetic dipole transitions}
The possibility of magnetic dipole transitions was raised by \cite{pachucki:11} who reported the corresponding transition moment 
$g(r)$ as a function of the internuclear distance $r$ and computed the corresponding emission probabilities within the $v$ = 1 $\rightarrow$ 0  transition. 
The emission probability, expressed in s$^{-1}$ is given by:
\begin{align}
W_{v'J' \rightarrow v"J"} &= \frac{4 \alpha \omega^3 }{3c^2}\cdot \frac{\alpha^2 a_0^2}{4} \cdot \left(\frac{m_\mathrm{e}}{M_\mathrm{P}}\right)^2 \cdot \frac{1}{(2J'+1)} \cdot\\\nonumber 
&\sum_{M'M"} \left|\left<J'M' f_{v',J'}(r) \left|g(r) \cdot  {\vec{\mathbf{J}}} \right| J"M" f_{v",J"}(r)\right>\right|_{CGS}^2
\end{align}
 \citep{sobelman:06}. Only $\Delta J = 0$ and $M' = M"$ transitions are allowed in this case from the ${\vec{\mathbf{J}}} $ matrix element:
\begin{equation}
\langle JM |{\bf{J}} | JM\rangle= \sqrt{J (J+1) (2J+1)}\,.
\end{equation}
Introducing the wavenumber $\sigma$ as previously and using the transition moment function $g(r)$
in atomic units we obtain:
\begin{equation}
W_{v'J \rightarrow v"J}= {8.00 \times 10^{-18}} \cdot  \sigma^3 \cdot J (J+1)  \cdot 
\left|\left< f_{v',J}(r) | g(r)|  f_{v",J}(r) \right>\right|^2\,.
\end{equation}
 
\subsection{Present computations}
 The emission probabilities involve both the transition wavenumbers and the radial integration of the corresponding matrix elements.
 We computed the matrix elements from the $f_{v,J}(r)$ solutions of the one dimension radial Schr{\"o}dinger equation corresponding to the \HH~ground state, by using the renormalized Numerov method \citep{johnson:77}:
 \begin{equation}
- \frac{\hbar^2}{2M} \frac{d^2f_{v,J}(r)}{dr^2} - \frac{\hbar^2}{2M} \frac{J (J+1)}{r^2}f_{v,J}(r) + V(r) f_{v,J}(r) =  E_{v,J} f_{v,J}(r),
\end{equation} 
where $M=M_\mathrm{P}/2$ is the nuclear reduced mass of \HH, $V(r)$ is the adiabatic potential 
function composed of the Born-Oppenheimer potential reported in \cite{Pachucki:10b}
and the adiabatic correction function presented in \cite{PK:14}.%to which non-adabatic corrections \citep{wolniewicz:95} have been appended. 

The $E_{v,J}$ eigenvalues correspond to the discrete rovibrational energies of \HH. We have verified that the computed emission probabilities are in excellent agreement with those reported previously in \cite{wolniewicz:98} and in \cite{pachucki:11}. Here, we extend the computations of the magnetic dipole transitions of \cite{pachucki:11} to all possible  $v'J - v"J$ transitions.

\section{Results and discussion}
The informations concerning  {\it{all}} possible transitions within the ground electronic state of \HH ~ (4712) are given  in electronic format \footnote {Full Table 2 is only available at the CDS via anonymous ftp to cdsarc.u-strasbg.fr (130.79.128.5) or via http://cdsarc.u-strasbg.fr/viz-bin/cat/J/A+A/630/A58. It can also be obtained on request to evelyne.roueff@obspm.fr}. Table \ref{tab:2} provides
the first rows  of the datafile.
In addition to the quantum numbers involved in the transitions, we display the transition wavenumbers $\sigma$ in cm$^{-1}$ and their theoretical estimated accuracy (see Section~\ref{Sec:lp}), the wavelengths in micron and the resulting accuracy, the electric quadrupole transition probability A$_{qu}$, the magnetic dipole transition probability A$_{ma}$, the sum  A = A$_{qu}$ + A$_{ma}$ giving the transition probability of the transition, the inverse of the total radiative lifetime $\tau$ of the upper energy level of the transition $A_{tot} = \sum_{l} A_{u \rightarrow l}$ in s$^{-1}$. We then report the energy terms of the upper level where the origin 0 is taken for the infinite separation of the two hydrogen atoms and the estimated accuracy in the same units. The next column gives the energy of the upper level expressed in Kelvin, when measured from the ground rovibrational state, as this value is used by astrophysicists when analysing the observed emission spectrum in order to derive excitation temperature, and the last column stands for the statistical weight of the upper level of the transition. We recall that the statistical weight of a particular level is given by $ (2J+1) \times g_I$ where $g_I$ = 1 for even values of $J$ (para levels) and $g_I$ = 3 for odd values of $J$  (ortho levels). 
{{\small
\begin{sidewaystable*}
\caption{List and properties of rovibrational transitions of \HH.
{\tiny{ (1) vibrational quantum number of the upper level;
(2) rotational quantum number of the upper level;
(3)  vibrational quantum number of the lower level;
(4) rotational quantum number of the lower level;
(5) wavenumber of the transition $v_u,J_u  \rightarrow  J_l,v_l$;
(6) uncertainty on the wavenumber;
(7)  wavelength of the  $v_u,J_u  \rightarrow  J_l,v_l$ transition;
(8)  uncertainty on the wavelength;  
(9) electric quadrupole transition emission probability;
(10) magnetic dipole  transition emission probability;
(11) full radiative transition emission probability;
(12) total radiative decay probability of the upper level;
(13) energy of the upper level measured from the dissociation limit of \HH;
(14) uncertainty on the energy;
(15) energy term of the upper level measured from the ground v=0, J=0 level, computed from the dissociation value of 36 118.0695 cm$^{-1}$ (present calculations);
(16) statistical weight of the upper level.
}}
}             % title of Table
\label{tab:2}      % is used to refer this table in the text
\centering                          % used for centering table
 \begin{tabular}{ c c c c c c c c c c c c c c c c }        % 16 centered columns (16 columns)
\hline\hline                 % inserts double horizontal lines
$v_u$& $J_u$& $v_l$& $J_l$& $\sigma$ &  $\Delta \sigma$ &$\lambda$  & $\Delta \lambda$ & A$_{qu}$ & A$_{ma}$ & A& A$_{tot}$& $E_u$& $\Delta E_u$& $E_u$& $g_u$ \\    % table heading 
   &  &   &   & {\tiny{cm$^{-1}$}}&   {\tiny{ cm$^{-1}$}}&  {\tiny{  $\mu$}} &  {\tiny{  $\mu$}}&  {\tiny{  s$^{-1}$ }}& {\tiny{  s$^{-1}$}} & {\tiny{  s$^{-1}$}} & {\tiny{  s$^{-1}$ }}& {\tiny{  cm$^{-1}$}}& {\tiny{  cm$^{-1}$}}& {\tiny{   K}}  & \\
\tiny{(1)}  &   \tiny{(2)}    &  \tiny{(3)}  & \tiny{(4)} & \tiny{(5)}  &(\tiny{6)} & \tiny{(7)} & \tiny{(8)} & \tiny{(9)} & \tiny{(10)}  & \tiny{(11)}  & \tiny{(12)} &\tiny{ (13)} & \tiny{(14)} & \tiny{(15) }& \tiny{(16)} \\
   \hline                        % inserts single horizontal line
{\tiny{0}} & {\tiny{2}} & {\tiny{0}}&  {\tiny{0}}& {\tiny{ 354.373130}} &{\tiny{ 3.7E-06}}  & {\tiny{  28.218843793}}& {\tiny{  2.9E-07}}& {\tiny{  2.943E-11}}  &  {\tiny{0.000E+00  } }& {\tiny{  2.943E-11 } } &  {\tiny{ 2.943E-11  }} &  {\tiny{  -35763.696396  }}&{\tiny{ 2.3E-04  }} &   {\tiny{    509.9 } }  &{\tiny{   5 } }\\
{\tiny{0}}   &{\tiny{3}}  & {\tiny{0}}  & {\tiny{1}}   &    {\tiny{ 587.032025}}  &{\tiny{6.1E-06 }}   & {\tiny{17.034845756}}  & {\tiny{1.8E-07 }} & {\tiny{  4.761E-10}}   & {\tiny{ 0.000E+00}}  &   {\tiny{4.761E-10}}  &  {\tiny{ 4.761E-10}}  &   {\tiny{ -35412.550688 }} &{\tiny{2.3E-04 }}  &   {\tiny{  1015.1 }}  & {\tiny{ 21}} \\
{\tiny{0}}  &{\tiny{4}} &  {\tiny{0}} &  {\tiny{2}}     &   {\tiny{814.424302}}  &{\tiny{8.4E-06}}   &  {\tiny{12.278611991}}  & {\tiny{1.3E-07}}  & {\tiny{  2.755E-09}}   & {\tiny{ 0.000E+00}}  &   {\tiny{2.755E-09}}   &  {\tiny{2.755E-09}}   &  {\tiny{ -34949.272093}}  &{\tiny{2.2E-04}}   &   {\tiny{  1681.6}}  &  {\tiny{9}} \\
{\tiny{0}} &  {\tiny{5}}   &{\tiny{0}}  & {\tiny{3}} &      {\tiny{1034.670685}}  &{\tiny{1.1E-05}}   &  {\tiny{ 9.664910918}}  & {\tiny{1.0E-07}}  &  {\tiny{ 9.836E-09}}   & {\tiny{ 0.000E+00}}  &   {\tiny{9.836E-09}}  &  {\tiny{ 9.836E-09}}  &   {\tiny{ -34377.880003}}  &{\tiny{2.2E-04}}   &   {\tiny{  2503.7}}  & {\tiny{ 33}} \\
{\tiny{0}}   &{\tiny{6}}  & {\tiny{0}} &  {\tiny{4}} &     {\tiny{ 1246.099547}}  &{\tiny{1.3E-05}}   & {\tiny{ 8.025041036}}  &{\tiny{ 8.4E-08}}   &{\tiny{  2.643E-08}}   & {\tiny{ 0.000E+00}}   & {\tiny{ 2.643E-08}}   &  {\tiny{2.643E-08}}   &   {\tiny{-33703.172546}} & {\tiny{2.1E-04}}   &   {\tiny{  3474.5}}  & {\tiny{ 13}} \\
{\tiny{0}}  &{\tiny{7}} &  {\tiny{0}} &  {\tiny{5}} &      {\tiny{1447.280936}}  &{\tiny{1.4E-05}}   &  {\tiny{ 6.909508549}}  &{\tiny{ 6.7E-08}}   & {\tiny{ 5.879E-08}}   & {\tiny{ 0.000E+00 }}  &  {\tiny{5.879E-08  }} &  {\tiny{5.879E-08}}   &  {\tiny{ -32930.599067}}  &{\tiny{2.0E-04}}   &    {\tiny{ 4586.1}}  &  {\tiny{ 45}} \\
{\tiny{0}}   &{\tiny{8 }} & {\tiny{0}}  & {\tiny{6}}  &   {\tiny{  1637.046000}}  &{\tiny{1.6E-05}}    & {\tiny{ 6.108563840}}  &{\tiny{ 6.0E-08}}   & {\tiny{ 1.142E-07}}   &  {\tiny{0.000E+00}}  &  {\tiny{ 1.142E-07}}   &  {\tiny{1.142E-07}}   &  {\tiny{ -32066.126547}} & {\tiny{2.0E-04}}   &    {\tiny{ 5829.8}}  &  {\tiny{17}} \\
{\tiny{0}}   &{\tiny{9}}   &{\tiny{0}} & {\tiny{7}} &      {\tiny{1814.492375 }} &{\tiny{1.7E-05}}    & {\tiny{5.511183259}}   &{\tiny{5.2E-08}}   &  {\tiny{2.001E-07}}   &  {\tiny{0.000E+00}}   &  {\tiny{2.001E-07}}    &{\tiny{ 2.001E-07}}   &  {\tiny{ -31116.106692}}  &{\tiny{1.9E-04 }}    &  {\tiny{ 7196.7}}  & {\tiny{ 57}} \\
{\tiny{0}}  &{\tiny{10}} &  {\tiny{0}}  & {\tiny{8}}  &   {\tiny{  1978.977263 }} &{\tiny{1.9E-05}}    & {\tiny{5.053115155}}  &{\tiny{ 4.9E-08 }}  &{\tiny{  3.236E-07 }}   & {\tiny{0.000E+00}}   & {\tiny{ 3.236E-07}}   & {\tiny{ 3.236E-07}}   &  {\tiny{ -30087.149284 }} &{\tiny{1.8E-04}}    &   {\tiny{ 8677.1}}  &  {\tiny{21}}  \\
  \hline                                   %inserts single line
\end{tabular}
\end{sidewaystable*}
}

A first comment concerns the wavelengths of the actual transitions taking place within the ground electronic state of \HH. As an example, \cite{pike:16} report the (2-1) S(27) transition of \HH\ at 2.1790 $\mu$ detection towards Herbig-Haro 7. The  wavelength is computed from the predictions given in \cite{dabrowski:84}. Former \citep{komasa:11} and present calculations show that the energy of the $v$=2, $J$=29 level  is not bound so that the transition is misidentified. Our present computations report a closebye transition at 2.1784 $\mu$ corresponding to  (8-5) S(14) with a very low emission probability of 2.52 10$^{-11}$ s$^{-1}$, which appears not realistic. 
 The \HH~quadrupole transition wavenumbers and corresponding emission probabilities are also displayed in the HITRAN database \citep{gordon:17}. We have checked the overall agreement between our computations  and those displayed in the HITRAN database. The transition wavenumbers values are restricted to four significant digits in HITRAN whereas the quoted uncertainty reported in the present work is variable and spans an interval between a few 10$^{-3}$ and a few 10$^{-6}$ cm$^{-1}$, so that some differences in the last digits can be obtained. The number of transitions reported in HITRAN2016 is not fully complete as a result of 
possible numerical difficulties linked to spline interpolations of the transition moments such as those arising for CO overtone transitions, as discussed in \cite{medvedev:16}. The accuracy of our potential function and electric quadrupole as well as magnetic dipole moments prevents the occurrence of such difficulties. 

A second comment concerns the relevance of the magnetic dipole contribution in the Q transitions of \HH ~which has been overlooked so far in astrophysical and plasma studies. 
In order to quantify their possible impact we have plotted both electric quadrupole A$_{qu}$ and magnetic dipole A$_{ma}$ for all Q transitions %over the full available spectral range in Figure \ref{fig:1}.
 by separating $\Delta v =1, \Delta v = 2$ and $\Delta v > 2$ transitions. The largest contribution of the magnetic dipole compared to electric quadrupole transition probabilities is obtained for $\Delta v = 1$ transitions and $\lambda$ above 3.5 $\mu$, which involve high $J$ rotational quantum numbers. This trend was already pointed out for the $v$ =1 $\to$  0 fundamental band Q transitions by  \cite{pachucki:11}. The differences can reach more than one order of magnitude.  The available spectroscopy measurements of line strengths from \cite{bragg:82} were restricted to low $J$ values ($J$ = 1, 2, 3 of the 1-0 band) and are insensitive to the magnetic dipole contribution. However, the state of the art techniques of intensity determination could allow to further check our derivations. In particular, the Q(6)-Q(10) transitions of the fundamental band  show already a 6 to 20 \% magnetic dipole contribution to the radiative transition probability, which could be challenged through experiments. 
%-------------------------------------------------------------
   \begin{figure}
   \centering
   \includegraphics[width=\hsize]{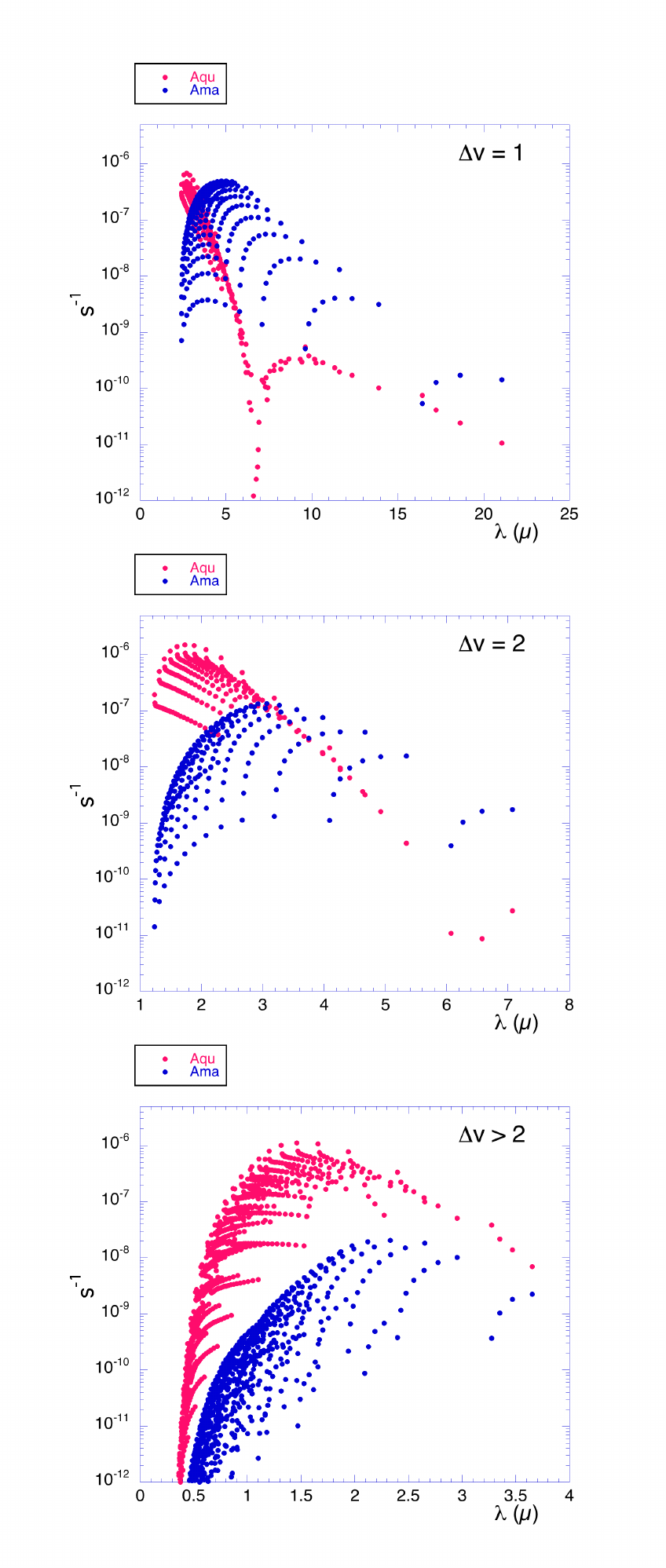}
      \caption{Electric quadupole A$_{qu}$ (red) and magnetic dipole A$_{ma}$  (blue) contributions to the transition probabilities in s$^{-1}$ for the Q branches of the \HH~rovibrational spectrum within the ground state for $\Delta v = 1$, $\Delta v = 2$,  $\Delta v > 2$.
              }
         \label{fig:1}
   \end{figure}
%
%-------------------------------------------------------------
Finally, we think that the present computations provide the most accurate \HH~ transition wavenumbers, wavelengths, emission probabilities which should be used in the analysis of high temperature plasma  and astrophysical conditions.

\begin{acknowledgements}
      We thank the referee for his(her) pertinent  suggestions which helped to improve the paper. Part of this work was supported by  the Programme National  de Physique et Chimie du Milieu Interstellaire  (PCMI) of CNRS/INSU with INC/INP co-funded by CEA and CNES.
      The computational part of this work was supported by NCN (Poland) grant 2017/25/B/ST4/01024 
      as well as by a computing grant from the Poznan Supercomputing and Networking Center.
 \end{acknowledgements}

\bibliographystyle{aa}
\bibliography{h2}

\begin{thebibliography}{32}
\expandafter\ifx\csname natexlab\endcsname\relax\def\natexlab#1{#1}\fi

\bibitem[{{Bragg} {et~al.}(1982){Bragg}, {Brault}, \& {Smith}}]{bragg:82}
{Bragg}, S.~L., {Brault}, J.~W., \& {Smith}, W.~H. 1982, \apj, 263, 999

\bibitem[{Campargue {et~al.}({2012})Campargue, Kassi, Pachucki, \&
  Komasa}]{Campargue:12}
Campargue, A., Kassi, S., Pachucki, K., \& Komasa, J. {2012}, Phys. Chem. Chem.
  Phys., {14}, {802}

\bibitem[{Caswell \& Lepage(1986)}]{Caswell:86}
Caswell, W.~E. \& Lepage, G.~P. 1986, Phys. Lett. B, 167, 437

\bibitem[{Czachorowski {et~al.}(2018)Czachorowski, Puchalski, Komasa, \&
  Pachucki}]{CPKP:18}
Czachorowski, P., Puchalski, M., Komasa, J., \& Pachucki, K. 2018, Phys. Rev.
  A, 98, 052506

\bibitem[{{Dabrowski}(1984)}]{dabrowski:84}
{Dabrowski}, I. 1984, Canadian Journal of Physics, 62, 1639

\bibitem[{{Geballe} {et~al.}(2017){Geballe}, {Burton}, \& {Pike}}]{geballe:17}
{Geballe}, T.~R., {Burton}, M.~G., \& {Pike}, R.~E. 2017, \apj, 837, 83

\bibitem[{{Gordon} {et~al.}(2017){Gordon}, {Rothman}, {Hill}, {Kochanov},
  {Tan}, {Bernath}, {Birk}, {Boudon}, {Campargue}, {Chance}, {Drouin}, {Flaud},
  {Gamache}, {Hodges}, {Jacquemart}, {Perevalov}, {Perrin}, {Shine}, {Smith},
  {Tennyson}, {Toon}, {Tran}, {Tyuterev}, {Barbe}, {Cs{\'a}sz{\'a}r}, {Devi},
  {Furtenbacher}, {Harrison}, {Hartmann}, {Jolly}, {Johnson}, {Karman},
  {Kleiner}, {Kyuberis}, {Loos}, {Lyulin}, {Massie}, {Mikhailenko},
  {Moazzen-Ahmadi}, {M{\"u}ller}, {Naumenko}, {Nikitin}, {Polyansky}, {Rey},
  {Rotger}, {Sharpe}, {Sung}, {Starikova}, {Tashkun}, {Auwera}, {Wagner},
  {Wilzewski}, {Wcis{\l}o}, {Yu}, \& {Zak}}]{gordon:17}
{Gordon}, I.~E., {Rothman}, L.~S., {Hill}, C., {et~al.} 2017, \jqsrt, 203, 3

\bibitem[{{Jennings} \& {Brault}(1983)}]{jennings:83}
{Jennings}, D.~E. \& {Brault}, J.~W. 1983, Journal of Molecular Spectroscopy,
  102, 265

\bibitem[{{Johnson}(1977)}]{johnson:77}
{Johnson}, B.~R. 1977, \jcp, 67, 4086

\bibitem[{{Kaplan} {et~al.}(2017){Kaplan}, {Dinerstein}, {Oh}, {Mace}, {Kim},
  {Sokal}, {Pavel}, {Lee}, {Pak}, {Park}, {Sok Oh}, \& {Jaffe}}]{kaplan:17}
{Kaplan}, K.~F., {Dinerstein}, H.~L., {Oh}, H., {et~al.} 2017, \apj, 838, 152

\bibitem[{{Ko{\l}os} \& {Wolniewicz}(1964)}]{kolos:64}
{Ko{\l}os}, W. \& {Wolniewicz}, L. 1964, \jcp, 41, 3674

\bibitem[{{Komasa} {et~al.}(2011){Komasa}, {Piszczatowski}, {Lach}, M.,
  {Jeziorski}, \& {Pachucki}}]{komasa:11}
{Komasa}, J., {Piszczatowski}, K., {Lach}, G., {et~al.} 2011, Journal of
  Chemical Theory and Computation, 7, 3105

\bibitem[{Komasa {et~al.}(2019)Komasa, Puchalski, Czachorowski, Lach, \&
  Pachucki}]{Komasa:19}
Komasa, J., Puchalski, M., Czachorowski, P., Lach, G., \& Pachucki, K. 2019,
  Submitted to Physical Review A

\bibitem[{{Le} {et~al.}(2017){Le}, {Pak}, {Kaplan}, {Mace}, {Lee}, {Pavel},
  {Jeong}, {Oh}, {Lee}, {Chun}, {Yuk}, {Pyo}, {Hwang}, {Kim}, {Park}, {Sok Oh},
  {Yu}, {Park}, {Minh}, \& {Jaffe}}]{le:17}
{Le}, H.~A.~N., {Pak}, S., {Kaplan}, K., {et~al.} 2017, \apj, 841, 13

\bibitem[{{Medvedev} {et~al.}(2016){Medvedev}, {Meshkov}, {Stolyarov},
  {Ushakov}, \& {Gordon}}]{medvedev:16}
{Medvedev}, E.~S., {Meshkov}, V.~V., {Stolyarov}, A.~V., {Ushakov}, V.~G., \&
  {Gordon}, I.~E. 2016, Journal of Molecular Spectroscopy, 330, 36

\bibitem[{{Oh} {et~al.}(2016){Oh}, {Pyo}, {Kaplan}, {Yuk}, {Park}, {Mace},
  {Park}, {Chun}, {Pak}, {Kim}, {Sok Oh}, {Jeong}, {Yu}, {Lee}, {Kim}, {Hwang},
  {Lee}, {Nguyen Le}, {Lee}, \& {Jaffe}}]{oh:16}
{Oh}, H., {Pyo}, T.-S., {Kaplan}, K., {et~al.} 2016, \apj, 833, 275

\bibitem[{Pachucki(2005)}]{Pachucki:05}
Pachucki, K. 2005, Phys. Rev. A, 71, 012503

\bibitem[{Pachucki(2010)}]{Pachucki:10b}
Pachucki, K. 2010, Phys. Rev. A, 82, 032509

\bibitem[{Pachucki \& Komasa(2008)}]{PK08b}
Pachucki, K. \& Komasa, J. 2008, J. Chem. Phys., 129, 034102

\bibitem[{{Pachucki} \& {Komasa}(2009)}]{pachucki:09}
{Pachucki}, K. \& {Komasa}, J. 2009, \jcp, 130, 164113

\bibitem[{Pachucki \& Komasa(2009)}]{PK:09}
Pachucki, K. \& Komasa, J. 2009, J. Chem. Phys., 130, 164113

\bibitem[{{Pachucki} \& {Komasa}(2011)}]{pachucki:11}
{Pachucki}, K. \& {Komasa}, J. 2011, Physical Review A, 83, 032501

\bibitem[{Pachucki \& Komasa(2014)}]{PK:14}
Pachucki, K. \& Komasa, J. 2014, J. Chem. Phys., 141, 224103

\bibitem[{Pachucki \& Komasa(2015)}]{PK:15}
Pachucki, K. \& Komasa, J. 2015, J. Chem. Phys., 143, 034111

\bibitem[{Pachucki \& Komasa(2018)}]{PK:18a}
Pachucki, K. \& Komasa, J. 2018, Phys. Chem. Chem. Phys., 20, 247

\bibitem[{{Pike} {et~al.}(2016){Pike}, {Geballe}, {Burton}, \&
  {Chrysostomou}}]{pike:16}
{Pike}, R.~E., {Geballe}, T.~R., {Burton}, M.~G., \& {Chrysostomou}, A. 2016,
  \apj, 822, 82

\bibitem[{Piszczatowski {et~al.}(2009)Piszczatowski, Lach, Przybytek, Komasa,
  Pachucki, \& Jeziorski}]{PLPKPJ09}
Piszczatowski, K., Lach, G., Przybytek, M., {et~al.} 2009, J. Chem. Theory
  Comput., 5, 3039

\bibitem[{Puchalski {et~al.}(2016)Puchalski, Komasa, Czachorowski, \&
  Pachucki}]{PKCP:16}
Puchalski, M., Komasa, J., Czachorowski, P., \& Pachucki, K. 2016, Phys. Rev.
  Lett., 117, 263002

\bibitem[{Puchalski {et~al.}(2017)Puchalski, Komasa, \& Pachucki}]{PKP:17}
Puchalski, M., Komasa, J., \& Pachucki, K. 2017, Phys. Rev. A, 95, 052506

\bibitem[{{Sobelman}(2006)}]{sobelman:06}
{Sobelman}, I.~I. 2006, {Theory of Atomic Spectra} (Alpha Science International
  Ltd)

\bibitem[{{Turner} {et~al.}(1977){Turner}, {Kirby-Docken}, \&
  {Dalgarno}}]{turner:77}
{Turner}, J., {Kirby-Docken}, K., \& {Dalgarno}, A. 1977, \apjs, 35, 281

\bibitem[{{Wolniewicz} {et~al.}(1998){Wolniewicz}, {Simbotin}, \&
  {Dalgarno}}]{wolniewicz:98}
{Wolniewicz}, L., {Simbotin}, I., \& {Dalgarno}, A. 1998, \apjs, 115, 293

\end{thebibliography}
  
\end{document}